\documentclass{article}
\usepackage{graphicx,color,amsmath}
\newcommand{\Slash}[1]{\ooalign{\hfil/\hfil\crcr$#1$}}
\newcommand{\pa}{\partial}
\newcommand{\nn}{\nonumber}
\newcommand{\psibar}{{\bar \psi}}
\newcommand{\ep}{\epsilon}
\newcommand{\bref}[1]{(\ref{#1})}
\newcommand{\cbar}{{\bar c}}

\begin{document}
\title{Quantum Master Equation for QED\\ in Exact Renormalization Group}
\author{Yuji Igarashi${}^a$,  Katsumi Itoh${}^a$ and Hidenori Sonoda${}^b$}

\maketitle
\begin{center}
${}^a$ Faculty of Education, Niigata University

${}^b$ Physics Department, Kobe University
\end{center}

\begin{abstract}

  Recently, one of us (H.S.) gave an explicit form of the
  Ward-Takahashi identity for the Wilson action of QED. We first
  rederive the identity using a functional method.  The identity makes
  it possible to realize the gauge symmetry even in the presence of a
  momentum cutoff. In the cutoff dependent realization, the abelian
  nature of the gauge symmetry is lost, breaking the nilpotency of the
  BRS transformation.  Using the Batalin-Vilkovisky formalism, we
  extend the Wilson action by including the antifield contributions.
  Then, the Ward-Takahashi identity for the Wilson action is lifted to
  a quantum master equation, and the modified BRS transformation
  regains nilpotency.  We also obtain a flow equation for the extended
  Wilson action.

\end{abstract}

\newpage

\section{Introduction}

One of the most important subjects in the exact renormalization group
(ERG) \cite{Wilson, Wegener, Polchinski} is to find a method to treat
gauge symmetry, which is naively incompatible with a momentum cutoff
$\Lambda$ introduced as a regularization (see ref.~\cite{Morris1} and
references therein for an alternative attempt to construct a gauge
invariant regularization scheme.).  Constraints on the induced symmetry
breaking terms are described by some identities for Green functions, so
called ``the broken Ward-Takahashi (WT) identities'' or ``the modified
Slavnov-Taylor (ST) identities''\cite{Becchi93,Ellwanger}.  These
identities are written either for the Wilson action S or its Legendre
transformed effective action $\Gamma$.  They have additional regulator dependent
terms which are absent in the standard WT or ST identities for the cut-off
removed theories. This is why they are called ``broken'' or
``modified'' identities.

Even if the standard realization of the symmetry no longer works, the
symmetry can be realized in a regularization dependent way.  This is
most conveniently done using the Batalin-Vilkovisky (BV) antifield
formalism\cite{BV,BV2}(see also ref.~\cite{Zinn-Justin}), which has been
recognized as the most general and powerful method for dealing with
symmetries.  Any local as well as global symmetries are described by the
quantum master equation (QME) in the BV formalism. It was shown
\cite{igarashi1} that if the QME holds for a cutoff-removed action
${\cal S}[\phi]$, this should also be the case for the Wilson action
$S[\Phi]$ with a finite value of momentum cutoff. Hence, at least
conceptually, the presence of the symmetry along ERG flow is
established.  This result may suggest that, for consideration of
symmetries, the use of the Wilson action $S$ is preferable to that of
the Legendre action $\Gamma$.

The above argument remains, however, at a formal level, and only little
has been known concerning how actions satisfying the QME looks like for
concrete cases, especially for gauge theories.\footnote{As for global
symmetries, see ref.~\cite{global}.  The RG flow equation for QED was
studied in ref.~\cite{Aoki-Gies} with some approximations.
Ref.~\cite{JP} discusses roles of the modified WT identity in relation
to numerical studies.} So far, only for the lattice chiral symmetry,
which is a prototype of the regularization dependent symmetry, it has
been shown that the QME contains the Ginsparg-Wilson relation and was
solved to give an action for self-interacting fermions \cite{igarashi2}.

Recently, an important result was obtained in formulating gauge symmetry
in terms of ERG: one of the present authors (H.S.) has derived the WT
identity \footnote{This WT identity is expected to carry the same
information as ``the broken WT identity'' for QED given in
ref.\cite{Bonini0, FreierWetterich} for the Legendre effective action
$\Gamma$. } $\Sigma_{\Phi}=0$ for the Wilsonian QED action $S[\Phi]$ and
its flow equation \cite{Sonoda, Sonoda-QED}.

From the WT identity, we obtain the BRS transformation $\delta$ that has
the following properties: 1) it depends on the Wilson action, and
therefore, is non-linear; 2) a non-trivial Jacobian factor associated
with $\delta$ is generated to cancel the change of the action; 3) it is
not nilpotent, $\delta^{2} \neq 0$.  The last property may be understood
as the absence of the abelian nature of the gauge algebra in the
presence of a momentum cutoff.

The main aim of the present paper is to describe how these results fit
in the formal argument in treating the gauge theory.  As pointed out
above, any symmetry, if it exists, can be described by the QME in the BV
antifield formalism. Actually we show that the WT identity
$\Sigma_{\Phi}=0$ for QED can be lifted to the QME $\Sigma_{\Phi,
\Phi^{*}}=0$.

In ref.~\cite{Sonoda}, the action $S[\Phi]$ that satisfies the WT
identity is obtained perturbatively.  Here we show how to construct the
action $S[\Phi,\Phi^*]$ that fulfills the QME, assuming that we have the
action satisfying the WT identity.  We call the action
$S[\Phi,\Phi^*]$ as the master action.

It is found that our master action, a formal solution to the QME, has
non-trivial antifield dependence: the infinite power series expansion
w.r.t. antifields takes the form of a Taylor expansion of the Dirac
fields, which corresponds to a shift of field variables in the Wilson
action.  As a byproduct of introducing antifields, we employ the
``quantum BRS transformation'' \cite{Lavrov}, $\delta_{Q}$, which is
assured to be nilpotent, $\delta_{Q}^{2}=0$, thanks to the QME.  Using
this, we show the BRS invariance of the Polchinski equation for our master
action.  We emphasize that the nilpotency is recovered in the BV
formalism even though the BRS transformation read off from the WT
identity is not nilpotent.

The rest of this paper is organized as follows. In the next section, we
give a general method for deriving the WT identity $\Sigma_{\Phi}=0$ for
a regularized theory, and apply it to QED to obtain the WT identity
derived in ref.~\cite{Sonoda}.  From the WT identity, we read off the
BRS transformation. In section 3, after a brief explanation of the
antifield formalism, we construct a master action for QED that satisfies
the QME $\Sigma_{\Phi, \Phi^{*}}=0$ starting from an action satisfying
the WT identity $\Sigma_{\Phi}=0$.  In the final section, the Polchinski
flow equation is given for our master action, and its BRS invarinace is
shown.

\section{Path integral derivation of WT identity\\ for cutoff QED}

We will derive the WT identity for the QED with a momentum cutoff
using the path integral formalism. Our derivation is based on the
method\footnote{In ref.\cite{Becchi93}, it is called ``the quantum
  action principle''.} which has already been discussed by several
authors \cite{Becchi93, Bonini, Morris2}.  The formalism given here may
be applicable to any theory with symmetry.  We will first discuss a
generic theory with a momentum cutoff, and then apply the results to
QED.

\subsection{WT identity for a cutoff theory}

The fields are denoted collectively by $\phi^{A}$. The index $A$
represents the Lorentz indices of tensor fields, the spinor indices of
the fermions, and/or indices distinguishing different types of 
fields.  The Grassmann parity for $\phi^{A}$ is expressed as
$\ep(\phi^{A})=\ep_{A}$, so that $\ep_{A} =0$ if the field $\phi^{A}$
is Grassmann even (bosonic) and $\ep_{A} =1$ if it is Grassmann odd
(fermionic).  The generating functional for this theory in the
presence of sources $J_A$ is given by
\begin{eqnarray}
{\cal Z}_{\phi}[J] = \int {\cal D} \phi 
\exp\left(-{\cal S}[\phi]+ J \cdot \phi
       \right ),
\label{part-func1}
\end{eqnarray}
where the action ${\cal S}$ is decomposed into the kinetic and
interaction terms
\begin{eqnarray}
{\cal S}[\phi] = \frac{1}{2}\phi \cdot D \cdot \phi + {\cal S}_{I}[\phi].
\label{micro-action}
\end{eqnarray}
In this paper we use the matrix notation in momentum space:
\begin{eqnarray}
J \cdot \phi&=& \int \frac{d^{d}p}{(2\pi)^d}J_{A}(-p)\phi^{A}(p), \nn\\
\phi \cdot  D  \cdot \phi &=& \int \frac{d^{d}p}{(2\pi)^d} 
\phi^{A}(-p) D_{AB}(p) \phi^{B}(p).
\label{cond-not}
\end{eqnarray}
We now introduce an IR momentum cutoff $\Lambda$ through a positive
function that behaves as
\begin{eqnarray}
 \quad K\Bigl(\frac{p}{\Lambda}\Bigr)\rightarrow \quad  \left\{
		\begin{array}{ll}
		 1 & (p^2 < \Lambda^2) \\
		  0 & (p^2 \rightarrow \infty )
		\end{array}
               \right.
\label{cutoff-func}
\end{eqnarray}
where the function goes to $0$ sufficiently rapidly as $p^{2}
\rightarrow \infty$.  For simplicity, we write the function as $K(p)$ in
the rest of the paper.
Using this function, we decompose the original fields $\phi^{A}$ with
the propagator $\left(D_{AB}(p)\right)^{-1}$ into two classes of
fields: the IR fields $\Phi^{A}$ with the propagator
$K(p)\left(D_{AB}(p)\right)^{-1}$, and the UV fields $\chi^{A}$ with
$(1-K(p))\left(D_{AB}(p)\right)^{-1}$.  To this end, we substitute a
gaussian integral over new fields $\theta^{A}$
\begin{eqnarray}
&{}&\hspace{-7mm}\int {\cal D} \theta \exp -\frac{1}{2}
\Bigl(\theta -  J(1-K)D^{-1}\Bigr) \cdot
\frac{D}{K(1-K)} \cdot 
\Bigl(\theta - (-)^{\epsilon(J)}D^{-1}(1-K)J\Bigr)
\nonumber\\
&{}&\hspace{8cm} = {\rm const}
\label{gauss}
\end{eqnarray} 
into the path-integral \bref{part-func1}, and introduce new variables
$\Phi$ and $\chi$ by
\begin{eqnarray}
\phi^{A} =  \Phi^{A} + \chi^{A},~~~~~~ \theta^{A}= (1-K)\Phi^{A}- K\chi^{A}.
\label{new-fields}
\end{eqnarray} 
Then, we obtain 
\begin{eqnarray}
{\cal Z}_{\phi}[J] &=& N_{J}\int {\cal D} \Phi{\cal D}\chi 
\exp-\biggl(\frac{1}{2}\Phi \cdot K^{-1}D \cdot \Phi - J \cdot
K^{-1}\Phi\nn\\
&{}& \hspace{3.1cm} +\frac{1}{2}\chi \cdot  (1-K)^{-1}D \cdot  \chi + 
{\cal S}_{I}[\Phi + \chi]\biggr)~,
\label{z-relation}
\end{eqnarray}
where
\begin{eqnarray}
N_{J} \equiv \exp \frac{1}{2}(-)^{\ep_{A}} J_{A}
 (1-K^{-1})\left(D^{-1}\right)^{AB} J_{B}.
\label{N-normalization}
\end{eqnarray} 
The Wilson action is given by 
\begin{equation}
S[\Phi] \equiv  \Phi \cdot K^{-1}D
\cdot \Phi /2 + S_{I}[\Phi]\label{wilsonaction}
\end{equation}
where $S_{I}[\Phi]$ is defined by 
\begin{eqnarray}
\exp -S_{I}[\Phi] \equiv \int{\cal D}\chi
\exp - \Bigl(\frac{1}{2} \chi \cdot (1-K)^{-1}D \cdot \chi + 
{\cal S}_{I}[\Phi + \chi]
\Bigr)~.
\label{Wilsonian}
\end{eqnarray} 
Note that the gaussian integral \bref{gauss} is chosen in such a way
that the UV fields $\chi^{A}$ do not couple to source terms, and hence
the Wilson action $S_{I}$ depends only on the IR fields
$\Phi^{A}$.\footnote{The construction of the Wilson action via similar
  techniques can be found in refs.~\cite{ Bonini, Wetterich,
    Morris3}. The idea of the decomposition of fields was also
  discussed in a non-local regularization scheme \cite{Paris}.}  The
partition function for $\Phi^{A}$
\begin{eqnarray}
Z_{\Phi}[J] = \int {\cal D} \Phi 
\exp\left(-S[\Phi]+K^{-1}J \cdot \Phi \right)
\label{part-2}
\end{eqnarray}
is related to that for $\phi$ by
\begin{eqnarray}
{\cal Z}_{\phi}[J] = N_{J} Z_{\Phi}[J].
\label{z-relation2}
\end{eqnarray}
This implies that the full generating functional $\mathcal{Z}_\phi$ can be
constructed from the Wilson action $S[\Phi]$.  In \bref{part-2}, note
that the source to the IR field $\Phi^{A}$ is multiplied by $K^{-1}$.
Therefore, the correlation functions in two theories are related as
\begin{eqnarray}
\left<\phi^{A_1} \cdots \phi^{A_N}\right>_{\phi}|_{J=0} 
= \left<(K^{-1}\Phi^{A_1})\cdots (K^{-1}\Phi^{A_N})\right>_{\Phi}|_{J=0}
~~~~(N \geq 3)~.
\label{correl-relation} 
\end{eqnarray} 
As for the two-point functions, there are extra contributions from the
factor $N_J$.

Now we consider how the IR cutoff affects the realization of symmetry.
Suppose the original gauge-fixed action ${\cal S}[\phi]$ is invariant
under the BRS transformation
\begin{eqnarray}
\phi^{A} \rightarrow \phi^{A \prime} =\phi^{A} +  \delta \phi^{A}~,
\qquad  \delta \phi^{A}= R^{A}[\phi]~\lambda~,
\end{eqnarray} 
where $\lambda$ is an anticommuting constant.  Hence,
\begin{eqnarray}
\delta {\cal S} = \frac{\pa^{r} {\cal S}}{\pa \phi^{A}} \delta \phi^{A} 
\equiv \Sigma_{\phi}~\lambda=0.
\end{eqnarray} 
Assuming the invariance of the functional measure ${\cal D} \phi$
\footnote{We assume the presence of BRS invariant regularization scheme
such as the dimensional regularization in order for the ${\cal
Z}_{\phi}$ theory to be well-defined. However, the knowledge of the ${\cal
Z}_{\phi}$ theory is only used as the boundary condition for the
$Z_{\Phi}$ theory at $\Lambda \to \infty$.
}, we
obtain the standard WT identity for ${\cal Z}_{\phi}$:
\begin{eqnarray}
\left<\Sigma_{\phi}\right>_{\phi, ~J}
&=& {\cal Z}_{\phi}^{-1}[J]  \int {\cal D} \phi~ 
J \cdot R[\phi] ~\exp\left(-{\cal S}[\phi]+J \cdot \phi\right)
\nn\\
&=& {\cal Z}_{\phi}^{-1}[J] ~\Bigl(J \cdot R [\pa^{l}_J ]~{\cal
  Z}_{\phi}[J]\Bigr) =0~.\label{WT1}
\end{eqnarray} 
Eq. (\ref{WT1}) may be rewritten as the WT identity for the
cutoff theory by using (\ref{z-relation2}),
\begin{eqnarray}
\left<\Sigma_{\phi}\right>_{\phi, ~J}
={\cal Z}_{\phi}^{-1}J \cdot R[\pa^{l}_J ]~{\cal Z}_{\phi}[J]
=Z_{\Phi}^{-1} \bigl(N_{J}^{-1}J \cdot R [\pa^{l}_J ] ~N_{J}Z_{\Phi}[J]\bigr)~.
\label{WT-Sigma-Phi}
\end{eqnarray} 
We expect that the last expression in the above can be written as the
expectation value of an operator in the cutoff theory: its vanishing
is a consequence of the symmetry of the original theory.  Therefore,
the operator is appropriately regarded as the ``WT operator''.  We
denote it by $\Sigma_{\Phi}$ so that
\begin{equation}
\left<\Sigma_{\Phi}\right>_{\Phi, ~K^{-1}J}=0
\end{equation}
In the next subsection, we obtain the WT operator explicitly for the
QED with a momentum cutoff.

\subsection{WT identity for the cutoff QED}

In addition to the gauge and Dirac fields $\{A_{\mu}, \psi,
\psibar\}$, we consider, for the BRS symmetry, the (non-interacting)
ghost and anti-ghost $\{c, \cbar\}$ as well as the auxiliary field
$B$.  Thus we have $\phi^{A}= \{A_{\mu}, B, c, \cbar, \psi, \psibar\}$
and the corresponding sources $J_{A}=\{J_{\mu}, J_{B}, J_{c},
J_{\cbar}, J_{\psi}, J_{\psibar}\}$.

The action is given as
\begin{eqnarray}
{\cal S}[\phi]&=& \frac{1}{2}\phi \cdot D \cdot \phi + {\cal
 S}_{I}[\phi]. \label{QED-action1}
\end{eqnarray}
The free part is 
\begin{eqnarray}
&&\frac{1}{2}\phi \cdot D \cdot \phi = \int_k \Bigl[\frac{1}{2} A_{\mu}(-k)
(k^2 \delta_{\mu\nu}-  k_{\mu}k_{\nu})A_{\nu}(k) + \cbar(-k)ik^{2}c(k)
\nn\\
&&\quad -B(-k)\bigl(ik_{\mu}A_{\mu}(k) + \frac{\xi}{2}B(k)\bigr)\Bigr]
+ \int_p \psibar(-p) (\Slash{p}+im) \psi(p),
\end{eqnarray}
where $\xi$ is the gauge parameter, and ${\cal S}_{I}[\phi]$ gives the
interaction part.  We assume that the above action is invariant under
the standard BRS transformation
\begin{eqnarray}
&&\delta A_{\mu}(k) = -i k_{\mu}~c(k), ~~~\delta \cbar (k) = i B(k), ~~
\delta c(k) = \delta B(k) =0~, \nn\\
&&\delta \psi(p) = -ie \int_{k} \psi(p-k)~c(k), 
~~~\delta \psibar (-p) = ie \int_{k}  \psibar(-p-k)~c(k).
\label{BRS2}
\end{eqnarray}
The source dependent normalization factor $N_J$ in (\ref{z-relation2})
can be calculated explicitly as
\begin{eqnarray}
\ln N_{J} &=& \frac{(-)^{\ep_{A}}}{2} 
J_{A} \Bigl(\frac{1-K}{K}\Bigr) \left(D^{-1}\right)^{AB} J_{B}\nn\\
&=& \int_k \Bigl(\frac{1-K}{K}\Bigr)(k) 
\Bigl\{
J_c(-k) \frac{-i}{k^2} J_{\bar c}(k) 
- J_B (-k) \frac{-ik_{\mu}}{k^2} J_{\mu}(k)\nn\\
&{}&~~~~~~- \frac{1}{2}J_{\mu}(-k) \frac{1}{k^2} 
\Bigl(\delta_{\mu\nu}-(1-\xi)\frac{k_{\mu}k_{\nu}}{k^2}\Bigr) 
J_{\nu}(k)
\Bigr\}
\nn\\
&{}&+\int_p \Bigl(\frac{1-K}{K}\Bigr)(p) 
J_{\psi}(-p) \frac{1}{\Slash{p}+i m} J_{\psibar}(p)
\label{N-J}
\end{eqnarray} 
The operator that appears in eq. (\ref{WT-Sigma-Phi}) takes the following
form for QED:
\begin{equation}
J \cdot R [{\pa^{l}_J}] = 
\left(J \cdot R [{\pa^{l}_J}] \right)_{\rm gauge} + 
\left(J \cdot R [{\pa^{l}_J}] \right)_{\rm matter}~,
\label{Slavnov1}
\end{equation}
where
\begin{eqnarray}
&&\left(J \cdot R [{\pa^{l}_J}] \right)_{\rm gauge}
= i \int_{k} \Bigl\{-k \cdot J(-k)\frac{\pa^{l}}{\pa J_{c}(-k)}
+ J_{\cbar}(-k)\frac{\pa^{l}}{\pa J_{B}(-k)}\Bigr\},\\
&&\left(J \cdot R [{\pa^{l}_J}] \right)_{\rm matter} = -ie \int_{p,~k}
\Big\{J_{\psi}(-p)~\frac{\pa^{l}}{\pa J_{\psi}(-p+k)} \nn\\
&&\hspace{4cm}- J_{\psibar}(p)~\frac{\pa^{l}}{\pa J_{\psibar}(p+k)}
\Big\} \frac{\pa^{l}}{\pa J_{c}(-k)}
\end{eqnarray} 

Let us now derive the WT-identity for the Wilson action
(\ref{wilsonaction}) for QED.  In the following, we use the same
notation for the IR fields as for the original fields: $\Phi^{A}=
\{A_{\mu}, B, c, \cbar, \psi, \psibar\}$.  The kinetic term is given
by
\begin{eqnarray}
\frac{1}{2}\Phi \cdot K^{-1}D \cdot \Phi
&=& \int_k K^{-1}(k) \Bigl[\frac{1}{2} A_{\mu}(-k)
(k^2 \delta_{\mu\nu}-  k_{\mu}k_{\nu})A_{\nu}(k) 
\nn\\
&~& ~~~~ + \cbar(-k)ik^{2}c(k) - B(-k)\bigl(ik \cdot A(k) +
 \frac{\xi}{2}B(k)\bigr)\Bigr]\nn\\
&~& + \int_p K^{-1}(p)\psibar(-p) (\Slash{p}+im) \psi(p) .
\label{kinetic-Wilson}
\end{eqnarray} 

It follows from \bref{WT-Sigma-Phi} that our central task for finding
$\Sigma_{\Phi}$ is to compute $Z_{\Phi}^{-1}N_{J}^{-1}J \cdot R
N_{J}Z_{\Phi}$.  It is easy to realize that the non-trivial
deformation from the standard WT identity has two origins: 1) the
normalization factor $N_{J}$, and 2) the scale factor $K^{-1}$ in the
source terms $K^{-1} J \cdot \Phi$ and in the kinetic terms $\Phi
\cdot K^{-1}D \cdot \Phi/2$.  Now, from \bref{WT-Sigma-Phi}, we have
\begin{eqnarray}
0 &=& Z_{\Phi}^{-1} \Bigl[ N_{J}^{-1}(J \cdot R) N_{J} \Bigr] Z_{\Phi}\nn\\
  &=& Z_{\Phi}^{-1} \Bigl[ (J \cdot R)_{\rm gauge}+N_{J}^{-1}(J \cdot
   R)_{\rm matter} N_{J} \Bigr] Z_{\Phi} .
\label{Phi relation}
\end{eqnarray}
The second line is a result of the fact, $(J \cdot R)_{\rm gauge}
N[J]=0$.  The matter sector which contains non-trivial contributions may
be written as follows:
\begin{eqnarray}
{\cal Z}_{\phi}^{-1}(J\cdot R)_{\rm matter}{\cal Z}_{\phi}&=& -ie \Biggl<
\int_{p,~k}\Bigl\{
\frac{J_{\psi}(-p)}{K(p)} U(-p, p-k)  \frac{J_{\psibar}(p-k)}{K(p-k)}\nn\\
&~&\hspace{-3cm}+ J_{\psi}(-p)~\frac{\pa^{l}}{\pa J_{\psi}(-p+k)}
- J_{\psibar}(p)~\frac{\pa^{l}}{\pa J_{\psibar}(p+k)} 
\Bigr\}c(k) 
\Biggr>_{\Phi,~K^{-1}J}
\label{matter JR}
\end{eqnarray}
where
\begin{eqnarray}
U(-p, p-k) \equiv \frac{1-K(p-k)}{\Slash{p}-\Slash{k}+im}K(p) - 
\frac{1-K(p)}{\Slash{p}+im}K(p-k) .
\label{matrix-U}
\end{eqnarray}
Using
\begin{eqnarray}
Z_{\Phi}^{-1}J_{A}Z_{\Phi} = \left<
K \frac{\pa^{r}S}{\pa \Phi^{A}}
\right>_{\Phi,~K^{-1}J} ,~~~~
Z_{\Phi}^{-1}\frac{\pa^{l}}{\pa J_{A}}Z_{\Phi} = \left<
K^{-1}\Phi^{A}\right>_{\Phi,~K^{-1}J},
\label{replacement-rules}
\end{eqnarray}
we obtain 
\begin{eqnarray}
\langle \Sigma_{\Phi} \rangle_{\Phi, K^{-1}J}=0
\label{WT identity for reg}
\end{eqnarray}
with
\begin{eqnarray}
&~&\hspace{-1.2cm}\Sigma_{\Phi} \equiv \int_{k}\Bigl\{\frac{\pa S}{\pa
  A_{\mu}(k)}(-ik_{\mu})c(k) + 
\frac{\pa^{r} S}{\pa \cbar(k)}iB(k) \Bigr\}\label{Sonoda-id}\\
&~&\hspace{-1.2cm}~~ -ie \int_{p,~k} \Bigl\{\frac{\pa^{r} S}{\pa
  \psi(p)}\frac{K(p)}{K(p-k)}\psi(p-k)
 -\frac{K(p)}{K(p+k)}\psibar (-p-k)
\frac{\pa^{l} S}{\pa \psibar(-p)} \Bigr\}c(k)\nn\\
&~&\hspace{-1.2cm}~ -ie \int_{p,~k} {\rm tr} \Bigl\{ \Bigl(
\frac{\pa^{l} S}{\pa \psibar(-p+k)} \frac{\pa^{r} S}{\pa \psi(p)}
- \frac{\pa^{l}\pa^{r} S}{\pa \psibar(-p+k)\pa \psi(p)}\Bigr)
U(-p,p-k)\Bigr\}c(k) .
\nn
\end{eqnarray}
From the identity (\ref{WT identity for reg}), we note that 
any correlation function with a $\Sigma_{\Phi}$ insertion vanishes,
\begin{eqnarray}
\left<\Sigma_{\Phi}~\Phi^{A_{1}}\Phi^{A_{2}}\cdots\Phi^{A_{N}}
\right>_{\Phi}|_{J=0}=0~.
\label{Sigma-id2}
\end{eqnarray}
Therefore, we obtain the operator identity $\Sigma_{\Phi}=0$, which is
the WT identity derived in refs. \cite{Sonoda, Sonoda-QED} for the
Wilson action of QED.

From eq. (\ref{Sonoda-id}), it is easy to realize that the WT identity
is nothing but the BRS invariance of the action under the standard BRS
transformation (\ref{BRS2}) as far as the gauge sector is concerned.
Since the gauge sector is free, this is quite natural.

Though the matter contribution to $\Sigma_\Phi$ is slightly
complicated, we will see presently that it also allows an
interpretation as a change of the action under some symmetry
transformation.  We may rewrite the matter contributions in
$\Sigma_{\Phi}$ as
\begin{eqnarray}
&{}& ie \int_{p,~k} \frac{\pa^{r} S}{\pa \psi(p)}~c(k)
\Bigl\{\frac{K(p)}{K(p-k)}\psi(p-k) - U(-p,p-k)
\frac{\pa^{l} S}{\pa \psibar(-p+k)} \Bigr\}\nn\\
&{}&-ie \int_{p,~k} \Bigl\{\frac{K(p)}{K(p+k)}\psibar(-p-k)\Bigr\}~c(k)  
\frac{\pa^{l} S}{\pa \psibar(-p)}\nn\\
&{}&+ie \int_{p,~k} {\rm tr}~\frac{\pa^{l}\pa^{r} S}{\pa \psibar(-p+k)\pa \psi(p)}U(-p,p-k)~c(k)~.
\label{rewrite}
\end{eqnarray}
From the first two lines of (\ref{rewrite}), we read off the BRS
transformation of the fermion.  Including the transformation for the
gauge sector, we find
\begin{eqnarray}
\delta A_{\mu}(k) &=& -i k_{\mu}~c(k), ~~~\delta \cbar (k) = i B(k), ~~
\delta c(k) = \delta B(k) =0 \nn\\
\delta \psi(p) &=& ie \int_{k} c(k)
\Bigl\{
\frac{K(p)}{K(p-k)}\psi(p-k) - U(-p,p-k)
\frac{\pa^{l} S}{\pa \psibar(-p+k)} 
\Bigr\}~,\nn\\
\delta \psibar (-p) &=& ie \int_{k} 
\Bigl\{
\frac{K(p)}{K(p+k)}\psibar (-p-k)
\Bigr\}c(k)~.
\label{BRS-tr-Wilson}
\end{eqnarray} 
With this transformation (\ref{BRS-tr-Wilson}), $\Sigma_{\Phi}$ is now
written as \cite{Sonoda-QED}
\begin{eqnarray}
\Sigma_{\Phi} = \frac{\pa^{r} S}{\pa \Phi^{A}} \delta \Phi^{A} 
+ ie~{\rm tr} \Bigl(\frac{\pa^{l}\pa^{r} S}{\pa \psibar\pa \psi}U\Bigr)c.
\label{Sigma-form}
\end{eqnarray}
The second term can be interpreted as the Jacobian factor associated with 
the BRS transformation \bref{BRS-tr-Wilson}.

We have three remarks on the BRS transformation
 \bref{BRS-tr-Wilson}:
(i) It depends on the Wilson action $S[\Phi]$, and
 therefore it is non-linear.
(ii) It is not unique: the non-linear contribution could appear both
 in $\delta \psi$ and $\delta \psibar$.
(iii) The nilpotency is lost on $\psi$, though it holds for other fields.

Obviously, the nilpotency is the most important property of the BRS
symmetry.  It is desirable to elevate (\ref{BRS-tr-Wilson}) to the one
with nilpotency.  In order to achieve this, we need to find out a way
to take care of the Jacobian factor appearing in (\ref{Sigma-form}).
This can be realized with the BV anti-field formalism.

\section{Antifield formalism and QME}

Let us first explain the antifield formalism briefly.
For each IR field $\Phi^{A}$, we introduce 
its antifield $\Phi_{A}^{*}$ with the opposite Grassmann parity, 
$\epsilon (\Phi_A^{*})= \epsilon (\Phi^{A}) +1$,
\begin{eqnarray}
\Phi^*_A &=& 
\{ A^*_{\mu},~B^*,~c^*,~{\bar c}^*,~\psi^*,~\psibar^* \}.
\label{Phi-prime}
\end{eqnarray}
The canonical structure of fields and their anti-fields is specified by
the anti-bracket. For any pair of operators, $X$ and $Y$, it is defined as
\begin{eqnarray}
(X,~Y) \equiv \frac{\partial^{r} X}{\partial
  \Phi^{A}}\frac{\partial^{l}Y}{\partial \Phi_{A}^*} 
- \frac{\partial^{r} X}{\partial
  \Phi^{*}_{A}}\frac{\partial^{l}Y}{\partial \Phi^{A}}~.  
\label{AB}
\end{eqnarray}

Consider a gauge theory with an action $S[\Phi,~\Phi^*]$ and
calculate the operator defined as
\begin{eqnarray}
\Sigma[\Phi,~\Phi^*] \equiv \frac{1}{2}(S, ~S) - \Delta S ,
\label{QME1}
\end{eqnarray}
where
\begin{eqnarray}
\Delta \equiv (-)^{\epsilon_A+1} 
\frac{\pa^r}{\pa \Phi^A}\frac{\pa^r}{\pa \Phi^*_A} .
\label{Delta}
\end{eqnarray}
The equation $\Sigma[\Phi,~\Phi^*]=0$ is the quantum master equation
of the BV formalism. The action satisfying the QME describes a gauge
invariant system.  The action satisfying $\Sigma[\Phi,~\Phi^*] = 0$ is
called a quantum master action, or simply a master action.  Later, we
denote a master action as $S_{M}[\Phi,~\Phi^*]$.

\vspace{2mm}

Our aim in this section is to construct a master action, by using the
WT identity \bref{Sonoda-id} for our Wilson action $S[\Phi]$.

In a standard gauge theory with a gauge-fixed action $S[\phi]$ and a
nilpotent BRS transformation $\delta \phi^A$, the master action is
$S[\phi] + \phi_{A}^{*} \delta \phi^{A}$, linear in anti-fields.  To
start with, let us try an extended action linear in the anti-fields
$\Phi_{A}^{*}$: $S_{\rm lin}[\Phi, \Phi^*]= S[\Phi] + \Phi_{A}^{*}
\delta \Phi^{A}$.  This action, however, does not satisfy the QME:~
$\Sigma[\Phi, \Phi^*] \propto c~c~\psi^{*} U~U$. To cancel this
contribution, one should add suitable terms $S_{\rm quad}[\Phi,
\Phi^*]$ quadratic in the anti-fields, and so on.  After several
trials, we realize that this expansion w.r.t. antifields is the Taylor
expansion of the action, where $\psibar$ is replaced by $\psibar
\rightarrow \psibar - ie \psi^{*}c~U$.

Let us assume this form for the master action and prove that it indeed
satisfies the QME.  Our master action is
\begin{eqnarray}
&~&\hspace{-7mm}S_M[\Phi, \Phi^*] = {S}[\Phi'] 
+ \int_k \Bigl( A_{\mu}^*(-k)(-i) k_{\mu} c(k) + i {\bar c}^*(-k) B(k)
\Bigr) 
\label{QMA}\\
&{}&\hspace{-3mm}+ ie \int_{p, k} \Bigl( 
\psi^*(-p) \frac{K(p)}{K(p-k)} c(k) \psi(p-k) + {\psibar}(-p-k)c(k)
\frac{K(p)}{K(p+k)}
{\psibar}^*(p)
\Bigr).
\nn
\end{eqnarray}
Here we have introduced the shifted field $\psibar'$
and $\Phi^{\prime A}$
\begin{eqnarray}
{\psibar}{'}(-p) &\equiv& {\psibar}(-p) - ie \int_k \psi^*(-p-k)
 c(k) ~U(-p-k, p)~,
\nn\\
\Phi^{\prime A} &=& \{ A_{\mu},~B,~c,~{\bar c},~\psi,~\psibar' \}~.
\label{Phi prime}
\end{eqnarray}
In eq. (\ref{QMA}), note that the second term in $\delta \psi$ of
(\ref{BRS-tr-Wilson}) is absorbed into $S[\Phi']$ due to the shift.

In proving the QME for $S_M$, it is important that the action
$S[\Phi]$ satisfies the WT identity.  For convenience, we rewrite the
identity for $S[\Phi']$ with the shifted fields (\ref{Phi prime}).  We
obtain
\begin{eqnarray}
&{}&\hspace{-5mm}
\int_{k}\Bigl\{\frac{\pa S[\Phi']}{\pa A_{\mu}(k)}(-ik_{\mu})c(k) + 
\frac{\pa^{r} S[\Phi']}{\pa \cbar(k)}iB(k) \Bigr\}\nn\\
&{}&\hspace{-5mm}
+ ie \int_{p, k} \frac{\pa^r {S}[\Phi']}{\pa \psi (p)} c(k)
\Bigl\{
\frac{K(p)}{K(p-k)}\psi(p-k) -U(-p,p-k)
\frac{\pa^{l} S[\Phi'] }{\pa \psibar(-p+k)} 
\Bigr\}
\nn\\
&{}&\hspace{-5mm} + ie \int_{p, k, l} \frac{\pa^r {S}[\Phi']}{\pa {\bar
 \psi}(-p)}\frac{K(p)}{K(p+k)}\nn\\
&{}&\hspace{3mm}\times \Bigl\{
{\bar \psi}(-p-k) - ie~\psi^*(-p-k-l) c(l)
U(-p-k-l, p+k)  
\Bigr\} c(k) \nn\\
&{}&\hspace{-5mm} + ie \int_{p, k} 
{\rm tr} \Bigl(\frac{\pa^l \pa^r {S}[\Phi']}{\pa \psibar(-p) \pa
\psi(p+k)}
U (-p-k, p) \Bigr) c(k) =0 .
\label{modified Sonoda's id}
\end{eqnarray}

Now it is straightforward to verify that the action \bref{QMA}
satisfies the QME.  Here we calculate the contributions to $(S_M,
S_M)/2$ from the matter sector.
\begin{eqnarray}
&{}&\int_p \frac{\pa^r S_M}{\pa {\bar \psi}(-p)}
\frac{\pa^l S_M}{\pa {\bar \psi}^*(p)}
= ie \int_{p,k} \frac{\pa^r {S}[\Phi']}{\pa {\bar \psi}(-p)} 
\frac{K(p)}{K(p+k)}
{\bar \psi}(-p-k)~c(k)~,
\label{cl part2}\\
&{}&\int_p \frac{\pa^r S_M}{\pa \psi (p)}
\frac{\pa^l S_M}{\pa \psi^*(-p)} 
\label{cl part3}\\
&&
= ie \int_{p,l}
\frac{\pa^r {S}[\Phi']}{\pa \psi (p)} c(l)
\Bigg(\frac{K(p)}{K(p-l)}\psi(p-l)- U(-p, p-l) \frac{\pa^l {S}[\Phi']}{\pa {\bar \psi}(-p+l)}
\Bigg)\nn\\
&{}&  \hspace{5mm} +  e^2 \int_{p,k,l} 
\frac{K(p+k+l)}{K(p+l)}
c(k)c(l) \psi^* (-p-k-l) U(-p-l,p)
\frac{\pa^l {S}[\Phi']}{\pa {\bar \psi}(-p)}~.
\nn
\end{eqnarray}
The quantum term may be calculated as
\begin{eqnarray}
\Delta_{\Phi} S_M = -ie \int_{p,k} 
{\rm tr}\Bigl(\frac{\pa^l \pa^r S_M}{\pa {\psibar}(-p) \pa \psi (p+k)}U(-p-k, p)\Bigr)c(k)~.
\label{quantum term}
\end{eqnarray}
Combining all the terms in \bref{cl part2}, \bref{cl part3} and 
\bref{quantum term}, we find 
\begin{eqnarray}
\Sigma [\Phi, \Phi^*] \equiv \frac{1}{2}(S_M, S_M)_{\Phi} -
 \Delta_{\Phi} S_M =0~,
\label{def of Sigma}
\end{eqnarray}
thanks to the identity \bref{modified Sonoda's id}.  Therefore the
action $S_M$ defined by (\ref{QMA}) is indeed a master action.  Note
that the same $e^2$ term appears in both (\ref{cl part3}) and
\bref{modified Sonoda's id}.

In summary, we have observed that the $(\partial S/\partial
\bar{\psi})(\partial S/\partial \psi)$ term of $\Sigma_\Phi$ is
absorbed into the classical part $(\partial S_M/\partial
\psi)(\partial S_M/\partial \psi^*)$ of the QME, corresponding to the
shift of $\bar{\psi}$.  Likewise, the $\partial \partial S/\partial
\bar{\psi} \partial \psi$ term of $\Sigma_\Phi$ turns into the
jacobian associated with the BRS transformation.  The shift of
$\bar{\psi}$ needed for constructing $S_M$ from $S$ now appears quite
natural.

In the antifield formalism, the ``quantum'' BRS transformation
\cite{Lavrov} is defined
by 
\begin{equation}
\delta_Q X \equiv (X,~S_{M})-\Delta X
\end{equation}
for any operator $X$.  For the fields in QED, it takes the following
form:
\begin{eqnarray}
&{}& \delta_Q A_{\mu}(k) = -i k_{\mu}~c(k), ~~~\delta_Q \cbar (k) = i B(k), ~~
\delta_Q c(k) = \delta_Q B(k) =0~,\nn\\
&{}& \delta_Q \psi(p) = ie \int_{k} c(k)
\Bigl\{
\frac{K(p)}{K(p-k)}\psi(p-k) - U(-p,p-k)
\frac{\pa^{l} S_{M}}{\pa \psibar(-p+k)} 
\Bigr\}~,\nn\\
&{}& \delta_Q \psibar (-p) = ie \int_{k} 
\Bigl\{
\frac{K(p)}{K(p+k)}\psibar (-p-k)
\Bigr\}c(k)~.
\label{BRS-tr-SM}
\end{eqnarray} 
This transformation has the same form as \bref{BRS-tr-Wilson}.  However,
the action on the r.h.s. of $\delta_Q \psi$ is now $S_{M}[\Phi,~\Phi^*]$,
and the BRS transformation has a non-trivial antifield dependence.  The
BRS transformation in the gauge sector is quite simple, while that of
the matter sector is rather complicated.

The quantum BRS transformation is nilpotent if and only if the QME
holds:
\begin{eqnarray}
\delta_{Q}^{2} X = \left(X,~ \Sigma [\Phi,~\Phi^{*}] \right) =0.
\label{delta-Q^2}
\end{eqnarray} 
In other words, the QME enables us to define the nilpotent BRS transformation.
This should be compared with the classical counterpart, $\delta X \equiv
\left(X, S_M\right)$ which does not 
vanish due to the lack of the Jacobian factor,
\begin{eqnarray}
\delta^{2} X =\frac{1}{2} \left(X,~ \left(S_{M},~S_{M}\right) \right)
 \neq 0.
\label{delta^2}
\end{eqnarray}

\section{Polchinski flow equation for the master action\\ and its BRS 
invariance}

In this section, we derive the Polchinski flow equation for our master
action and show its BRS invariance. 

Let us begin with the well-known generic result on the Polchinski flow equation 
for the Wilson action $S[\Phi]$ without antifields. It is given by 
\begin{eqnarray}
&{}&\hspace{-7mm}\pa_t S[\Phi] = - \int_p \Phi^A(p) \bigl(K^{-1} {\dot K}\bigr)(p) 
\frac{\pa^l {S}}{\pa \Phi^A(p)}\label{Polchinski1}\\
&{}&\hspace{-7mm}+\frac{1}{2}\int_p \Bigl[
\frac{\pa^r {S}}{\pa \Phi^A(p)}
\bigl({\dot K} D^{-1}(p)\bigr)^{AB}
\frac{\pa^l {S}}{\pa \Phi^B(-p)}
-(-)^{\epsilon_A}  \bigl({\dot K} D^{-1}(p)\bigr)^{AB}
\frac{\pa^l \pa^r {S}}{\pa \Phi^B(-p) \pa \Phi^A(p)}
\Bigr]
\nn
\end{eqnarray}
up to terms independent of fields. Here, we use a dimensionless 
parameter $t =\log(\Lambda/\mu)$ and ${\dot K} = \pa_{t} K$.  

The flow equation for our master action $S_{M}[\Phi,\Phi^*]$ of QED can
be obtained through a straightforward calculation. From the
definition \bref{QMA}, we have
\begin{eqnarray}
&{}&\hspace{-8mm}\pa_t S_M [\Phi,\Phi^*] = \pa_t S[\Phi]
\bigl|_{\Phi=\Phi^{\prime}} 
-ie \int_{p,k} \psi^*(-p-k) c(k) 
\pa_t U(-p-k, p)
%
\frac{\pa^l S[\Phi^{\prime}]}{\pa \psibar(-p)}
\label{pa t Sm 1}\\
&{}&\hspace{-5mm} +\int_{p, k} c(k)\Bigl[ \psi^*(-p) \pa_t 
\Bigl(\frac{K(p)}{K(p-k)}\Bigr)  \psi(p-k)- \psibar(-p-k) \pa_t
\Bigl(\frac{K(p)}{K(p+k)}\Bigr)  \psibar^*(p) \Bigr].
\nn
\end{eqnarray}
In replacing $S[\Phi']$ by $S_{M}[\Phi,\Phi^*]$, one takes account of
the following points specific to the abelian nature of QED: (1) the
ghost is a free field, and the BRS transformation for the gauge and
ghost sector $\{A_{\mu},~B,~c,~{\bar c}\}$ is cutoff independent; (2)
the shift of the fermionic field $\psibar \rightarrow \psibar -ie
c~\psi^{*}~ U$ generates a non-trivial antifield dependence in the
flow equation. We also note the following identity for the matrix $U$:
\begin{eqnarray}
\pa_t U(-p-k, p) &=& \Bigl(
\frac{{\dot K}(p+k)}{K(p+k)}+\frac{{\dot K}(p)}{K(p)}\Bigr)U(-p-k, p)\nn\\
&{}&\hspace{-1cm}+\frac{{\dot K}(p+k)}{K(p+k)}K(p)
\frac{1}{\Slash{p}+\Slash{k}+im} 
-\frac{{\dot K}(p)}{K(p)}K(p+k)\frac{1}{\Slash{p}+im}.
\label{partial U}
\end{eqnarray}
Then, putting altogether, we obtain
\begin{eqnarray}
&{}&\pa_t S_M [\Phi,\Phi^*]\nn\\
&{}&= \frac{1}{2} \int_p \frac{{\dot K}(p)}{p^2} \Bigl(
\delta_{\mu\nu}-(1-\xi)\frac{p_{\mu}p_{\nu}}{p^2} \Bigr)
\Bigl(
\frac{\pa S_M}{\pa A_{\mu}(p)}\cdot \frac{\pa S_M}{\pa A_{\nu}(-p)}
- \frac{\pa^2 S_M}{\pa A_{\mu}(p) \pa A_{\nu}(-p)}
\Bigr)
\nn\\
&{}&-\int_p \frac{{\dot K} (p)}{K(p)}
\Bigl[
A_{\mu}(p) \Bigl(
\delta_{\mu\nu}-\frac{p_{\mu}p_{\nu}}{p^2}
\Bigr)
\frac{\pa S_M}{\pa A_{\nu}(p)}
+\frac{i p_{\nu}}{p^2}
\xi B(p)
\frac{\pa S_M}{\pa A_{\nu}(p)} \Bigl]
\nn\\
&{}& +\int_p \frac{{\dot K} (p)}{K^2(p)}
\Bigl[B(-p)\bigl(i p \cdot A(p) + \xi B(p)\bigr)
- {i p^2} {\bar c}(-p) c(p) \Bigr]
\nn\\
&{}&+ \int_p {\dot K}(p) \Bigl[
\frac{\pa^r S_M}{\pa \psi (p)} \frac{1}{\Slash{p}+im}
\frac{\pa^l S_M}{\pa \psibar(-p)}
+ {\rm tr} \Bigl(\frac{1}{\Slash{p}+im} \cdot \frac{\pa^l \pa^r S_M}{\pa
\psibar (-p) \pa \psi (p)}
\Bigr)
\Bigr]
\nn\\
&{}&- \int_p \frac{{\dot K}(p)}{K(p)}
\Bigl(
\psibar (-p) \frac{\pa^l S_M}{\pa \psibar (-p)}
+ \frac{\pa^r S_M}{\pa \psi (p)} \psi (p)
-\psi^* (-p) \frac{\pa^l {S_M}}{\pa \psi^* (-p)}
- \frac{\pa^r {S_M}}{\pa \psibar^* (p)} \psibar^* (p)
\Bigr)
\nn\\
&{}&- i e \int_{p,k} \frac{{\dot K}(p)}{K(p)}K(p-k) c(k) 
\label{flow-SM}\nn\\
&{}&~~~~~~~\times \Bigl(
\psi^*(-p)
\frac{1}{\Slash{p}+im}
\frac{\pa^l {S_M}}{\pa \psibar(-p+k)}
-\frac{\pa^r {S_M}}{\pa \psi(p)}
\frac{1}{\Slash{p}+ im}  \psibar^*(p-k) \Bigr) .
\end{eqnarray}
Thanks to the abelian nature of the theory, no antifields appear in the
gauge and ghost sector.   The fermionic sector has explicit antifield
dependence.

Let us discuss the BRS invariance of the flow equation \bref{flow-SM}.
For the RG flow of the WT operator, we obtain the relation,
\begin{eqnarray}
  \pa_{t}~ \Sigma [\Phi,\Phi^{*}] = (\partial_t S_M ,~S_{M}) - \Delta
  \partial_t S_M = \delta_Q ~ \partial_t S_M =0,
\label{WT-flow}
\end{eqnarray} 
which implies that the flow itself should be written as a quantum BRS transform
of something.
Actually, up to the QME, we find \cite{igarashi1} 
\begin{eqnarray}
\partial_t S_M = - \delta_Q G,
\label{delta-G}
\end{eqnarray}
where $G$ is the generator of a canonical transformation
\begin{equation}
G = G_1 + G_2 + G_3
\end{equation}
that has three parts:
\begin{eqnarray}
G_1 &\equiv& \int_k A_{\mu}^*(-k) \Bigl[
\frac{1}{2} \frac{{\dot K}(k)}{k^2}
\Bigl(\delta_{\mu\nu}-(1-\xi)\frac{k_{\mu} k_{\nu}}{k^2}\Bigr)
\frac{\pa S_M}{\pa A_{\nu}(-k)}\nn\\
&{}&\qquad  + \frac{{\dot K}(k)}{K(k)}\frac{ik_{\mu}}{k^2} 
\Bigl(i k \cdot A(k) - \xi B(k)\Bigr)
\Bigr] ,\\
G_2 &\equiv& - \int_k \frac{{\dot K}(k)}{K(k)} 
\Bigl[A_{\mu}^{*}(-k) A_{\mu}(k) + B^*(-k)B(k) 
\label{canonical-tr-flow}\nn\\
&{}&~~~~~~+ {\bar c}^*(-k) {\bar c}(k) + \psi^*(-k) \psi(k)
+ {\bar \psi}(k) {\bar \psi}^*(-k)
\Bigr] ,\\
G_3 &\equiv& \hspace{-2mm}\int_p \psi^*(-p) \frac{{\dot K}(p)}{\Slash{p}+i m}
\Bigl[ \frac{\pa^l S_M}{\pa \psibar(-p)} 
-\frac{i e}{K(p)}\int_k c(k) K(p-k) \psibar^*(p-k)
\Bigr] .
\end{eqnarray}

\section{Summary and Discussion}

In this paper, we have rederived the WT identity for the Wilson action
of QED using a functional method and shown that it can be lifted to a
QME in the BV antifield formalism.  The master action, our formal
solution to the QME, generically has non-linear but simple anti-field
dependence which appears merely as a shift of field variables.  We
have also found that the master action is not unique, and that it can
be deformed by canonical transformations in the space of fields and
antifields.  No deformation can remove the non-linear anti-field
dependence in the master action.  We believe that the non-linear
anti-field dependence is an inherent feature of any local symmetries
in cutoff field theories.

We have also derived an extended flow equation for the master action.
Since the master action is determined up to canonical transformations,
the flow equation is not unique, and can be modified by canonical
transformations.

A pair of fundamental equations, the WT identity $\Sigma_{\Phi}=0$ and
the Polchinski equation $\partial_{t} S[\Phi]-{\cal F}[\Phi]=0$, can
be interpreted as a gauge fixed version of the QME and extended flow
equation:
\[
 \begin{array}{r@{~=~}l}
 \Sigma_{\Phi} & \Sigma[\Phi,\Phi^{*}]|_{\Phi^{*} \to 0}
  =0\\
 \partial_{t} S[\Phi]-{\cal F}[\Phi] & (\partial_{t} S_{M}[\Phi,\Phi^{*}]
  -{\cal F}[\Phi, \Phi^{*}])|_{\Phi^{*} \to 0} =0
 \end{array}
\]
It should be emphasized that the QME plays a crucial role not only in
constructing a nilpotent BRS transformation, but also in showing the
BRS invariance of the extended flow equation.  These properties imply
that the exact gauge symmetry does exist in the Wilson action of QED
despite the presence of a finite momentum cutoff.

A perturbative solution to the WT identity $\Sigma_{\Phi}=0$ and the
Polchinski equation $\partial_{t} S[\Phi]-{\cal F}[\Phi]=0$ has been
obtained in refs.~\cite{Sonoda,Sonoda-QED}.  It is
straightforward to to find the corresponding perturbative solution to
the QME: $\Sigma[\Phi,\Phi^{*}]=0$ and the extended flow equation:
$\partial_{t} S_{M}[\Phi,\Phi^{*}] -{\cal F}[\Phi, \Phi^{*}]=0$.

\vspace{5mm}

\noindent
{\bf Acknowledgments}

\noindent This work is supported in part by the Grants-in-Aid for
Scientific Research No. 13135209, 15540262 and 17540242 from the Japan
Society for the Promotion of Science.


\end{document}